\documentclass[aps,prb,reprint,showpacs,longbibliography,superscriptaddress]{revtex4-1}
\usepackage{graphicx,color,bm}
\usepackage{amsbsy,amssymb,amsmath,bm}

\usepackage{graphicx}
\usepackage{epstopdf}
\usepackage{latexsym}
\usepackage{subfigure}
\usepackage{color}
\usepackage{hyperref}

\newcommand{\KCB}{{K$_2$CuSO$_4$Br$_2$}}
\newcommand{\ccc}{{Cs$_2$CuCl$_4$}}
\newcommand{\be}{\begin{equation}}
\newcommand{\ee}{\end{equation}}
\newcommand{\bea}{\begin{eqnarray}}
\newcommand{\eea}{\end{eqnarray}}

\begin{document}

\title{Electron spin resonance in a model $S=1/2$ chain antiferromagnet with a uniform Dzyaloshinskii--Moriya
interaction}

\author{A.~I.~Smirnov}
\affiliation{P.~L.~Kapitza Institute for Physical Problems, RAS,
119334 Moscow, Russia}
\homepage{http://kapitza.ras.ru/rgroups/esrgroup/Welcome.html}

\author{T.~A.~Soldatov}
\affiliation{P.~L.~Kapitza Institute for Physical Problems, RAS, 119334 Moscow, Russia} \affiliation{{Moscow
Institute for Physics and Technology, 141700 Dolgoprudny, Russia}}

\author{{K.~Yu.~Povarov}}
\altaffiliation[Previous address: ]{P. L. Kapitza Institute for
Physical Problems RAS} \affiliation{Neutron Scattering and
Magnetism, Laboratory for Solid State Physics, ETH Z\"{u}rich,
Switzerland} \homepage{http://www.neutron.ethz.ch/}

\author{M. H\"alg }
\affiliation{Neutron Scattering and Magnetism, Laboratory for Solid State Physics, ETH Z\"{u}rich, Switzerland}

\author{W. E. A. Lorenz}
\affiliation{Neutron Scattering and Magnetism, Laboratory for Solid State Physics, ETH Z\"{u}rich, Switzerland}

\author{A. Zheludev}
\affiliation{Neutron Scattering and Magnetism, Laboratory for Solid
State Physics, ETH Z\"{u}rich, Switzerland}

\date{\today}

\begin{abstract}
The electron spin resonance spectrum of a quasi 1D $S=1/2$ antiferromagnet \KCB\ was found to demonstrate an
energy gap and a doublet of resonance lines in a wide temperature range between the Curie--Weiss  and Ne\`{e}l
temperatures.  This type of magnetic resonance absorption corresponds well to the two-spinon continuum of
excitations in $S=1/2$ antiferromagnetic spin chain with a uniform Dzyaloshinskii--Moriya interaction between
the magnetic ions. A resonance mode of paramagnetic defects demonstrating strongly anisotropic behavior due to
interaction with spinon excitations in the main matrix is also observed.
\end{abstract}

\maketitle

\section{Introduction}\label{Introduction}

One-dimensional $S=1/2$ antiferromagnets are intensively studied because of the numerous collective quantum
effects, including the spin-liquid behavior at zero temperature and fractionalized spin excitations
(spinons),\cite{FaddeevTakhtajan} revealing itself in a form of two-particle
continuum.\cite{KCuF3,LakeMultispinon} In applied magnetic field this continuum splits into transverse and
longitudinal branches and develops a fine structure.\cite{Mueller,DenderIC} Typically, this fine structure is
undetectable by the electron spin resonance (ESR) experiments probing strictly $q=0$ excitations. However, some
additional perturbations may dramatically modify the long-wavelength response of the spin
chain.\cite{OshikawaAffleck} We address now to a fine structure of the two-spinon continuum, appearing at $q=0$
in spin chains with the so-called uniform Dzyaloshinskii--Moriya (DM) interaction.

A feature of the {\it uniform} DM interaction is the parallel orientation of characteristic vectors ${\bf D}$
for all bonds within a chain. This pattern is fundamentally different from that of a conventional {\it
staggered} DM interaction, when vectors ${\bf D}$ compose an alternating antiparallel structure. The staggered
DM interaction is known to induce a canting of magnetic sublattices in a classical antiferromagnet, resulting in
a weak ferromagnetic moment. \cite{Dzyaloshinsky,Moriya} The uniform interaction would stabilize a spiral spin
structure in a classical chain. For a quantum spin chain, which has a disordered ground state, a uniform DM
interaction modifies the spectrum of excitations.\cite{StarykhESR,Povarov2011,Affleck2011}  In particular, this
causes a shift of the spinon continuum in momentum space by a specific wavevector $q_{\text{DM}}={D}/Ja$ (here
$J$ is the exchange integral and $a$ - interspin distance). As a result, in a magnetic field ${\bf
H}\parallel{\bf D}$ the ESR line splits into a doublet. The frequencies of the doublet components are at the
upper and lower boundaries of the initial (i.e. unshifted) continuum of transversal spin fluctuations at the
wavevector $q_{\text{DM}}$.\cite{StarykhESR,Affleck2011} Thus, the doublet is marking the width of the continuum
at this particular wavevector.

\begin{figure}[t]
\centering
\includegraphics[width=0.4\textwidth]{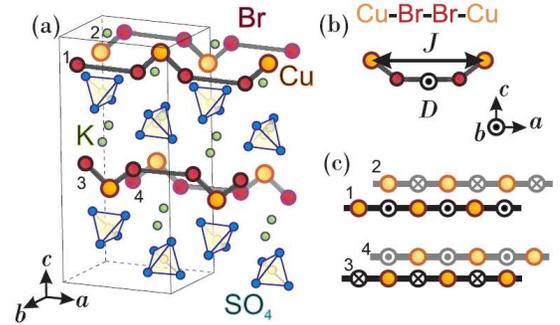}
 \caption{(Color online) (a) Schematic representation of \KCB\ crystal structure. (b) Magnetic interactions
 between the Cu$^{2+}$ ions as given in Hamiltonian~(\ref{Ham}). Vector ${\bf D}$ presents the symmetry allowed parameter of Dzyaloshinskii--Moriya
 interaction. (c) Two sets of nonequivalent spin chains with
 the opposite direction of DM vector.
 }\label{structure}
\end{figure}

 At the same time, ESR signal should not split at ${\bf H}\perp{\bf D}$. Another
consequence of the spectrum modification is a gap of the ESR absorption in zero field. This case should not be
confused with the gapped ``breather'' mode arising in nonzero field in spin chains with staggered DM, as
described in Refs.~\onlinecite{OshikawaAffleck,AjiroReview}. The described modifications stem from the
non-equivalence of spinons, propagating in different directions along the spin chain under the action of uniform
DM interaction and they are the consequence of the fractionalized nature of
excitations.\cite{StarykhESR,Affleck2011} Emerging fine structure enables one, in particular, to measure the
width of the continuum of transverse spin oscillations near the origin of the Brillouin zone by means of ESR.
The manifestation of the spinon continuum in ESR response was indeed observed in \ccc\
--- a frustrated $S=1/2$ quasi-2D distorted
triangular lattice antiferromagnet.\cite{Povarov2011,Smirnov2012,SmirnovPRB2015} The observed 1D behavior in a
nominally 2D system was associated with the effective decoupling of spin chains due to the frustration of
interchain bonds within the triangular lattice.

Recently,  a true  spin $S=1/2$ antiferromagnetic chain compound with uniform Dzyaloshinskii--Moriya interaction
\KCB\ was synthesized and studied by H\"{a}lg \emph{et al.}\cite{PovarovPRB2014} The structure of this
orthorhombic Pnma mineral is shown in Fig.~\ref{structure}. The magnetic $S=1/2$ Cu$^{2+}$ ions are linked by
the two-bromine superexchange pathways in chains, running along the $a$ axis. The value of the intrachain
exchange integral $J=20.5$~K is 600 times larger than the interchain exchange $J'$. Thus, the correspondence of
this material to the 1D model is much more exact  than for the above mentioned \ccc. Like other low-dimensional
magnets, \KCB\ does not order magnetically at cooling far below the Curie--Weiss temperature. It orders
antiferromagnetically at $T_{N}\simeq70$~mK, while the Curie--Weiss temperature is about 20 K, thus
demonstrating a very good correspondence to 1D model. The model Hamiltonian of a spin chain including Heisenberg
exchange and uniform Dzyaloshinskii--Moriya interactions may be written as:
\begin{equation}
{\mathcal{H}} = \sum\limits_{i} \left( J{\bf S}_i{\bf S}_{i+1} +{\bf D}\cdot{\bf S}_i \times {\bf S}_{i+1}
\right) \ . % + \mathcal{H_{\delta}}.
\label{Ham}
\end{equation}

Here  $J$ is the intrachain exchange integral and vector ${\bf D}$ is the parameter of DM interaction. For \KCB\
the DM vectors in adjacent chains are of the same absolute value but are directed along opposite directions,
being parallel to $b$-axis, as shown by the symmetry analysis.\cite{PovarovPRB2014} This arrangement is simpler
than in \ccc, where a few non-parallel DM vector directions exist in adjacent chains. The simple collinear
structure of DM vectors in this compound enables one to adjust the magnetic field parallel to all DM vectors,
which was impossible for \ccc . The observation of the splitting of the ESR line in \KCB\ at the frequency
$\nu=26.86$~GHz at cooling below 20 K was briefly reported in Ref.~\onlinecite{PovarovPRB2014}. Now we describe
the investigation of the ESR spectrum of this compound in a wide frequency and temperature range. Our
measurements cover the sub-gap and over-gap frequency domains, reconstructing the entire frequency-field
dependence of ESR absorption. The anisotropy of the effect is also checked to be consistent with the
spinon-based theory.\cite{StarykhESR} Besides, we follow the temperature dependence of the gap and of the
splitting of the doublet at cooling down to the temperature of $0.45$~K corresponding to the energy of
Dzyaloshinskii--Moriya interaction.

\section{Experimental details}\label{Experiment}

Experiments were performed using a set of ESR spectrometers, combined with a superconducting 12~T solenoid and a
$^3$He cryostat, providing temperature down to~$0.45$~K. A standard of 2,2-diphenyl-1-picrylhydrazyl (known as
DPPH) was employed as a $g=2.00$ marker for the magnetic field. Backward wave oscillators, Gann diodes and
klystrons were microwave sources, covering the range $0.5-250$~GHz. The microwave units with cylindrical,
rectangular, cut-ring and spiral resonators were used for recording the resonance absorption of microwaves.  In
case of a properly tuned resonator at frequency $\nu$ we observe the diminishing of the transmission,
proportional to the imaginary part of the uniform susceptibility of the sample $\chi''(\nu,0)$. The ESR line of
a conventional paramagnet, recorded in this case as a dependence of the transmission on the external field,
should have a Lorentzian shape. Unfortunately, for frequencies above 200~GHz the spectrum of eigenfrequencies of
the cavity is too dense and proper tuning is difficult, therefore a distortion of the ESR line is possible. The
presence of this parasitic effect was checked by recording  test ESR lines at higher temperatures in the
paramagnetic phase. In case of a parasitic distortion an error for the resonance field of about a half of the
resonance width was inserted.

%Fig2
\begin{figure}[t]
\centering
\includegraphics[width=0.5\textwidth]{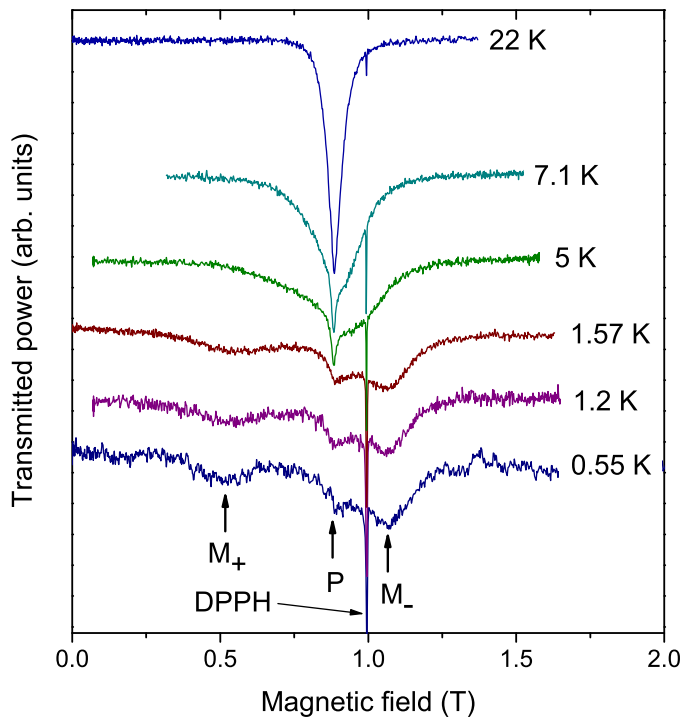}
 \caption{(Color online) Temperature evolution of the $27.83$~GHz ESR line of sample A  at ${\bf H}\parallel
 b$. $M_{+}$ and $M_{-}$ label the components of the doublet, $P$ is a paramagnetic line associated with residual defects.
 A narrow line at 0.994~T is the DPPH label.}\label{Tevol27GHzHb}
\end{figure}

%Fig3
\begin{figure}[t]
\centering
\includegraphics[width=0.5\textwidth]{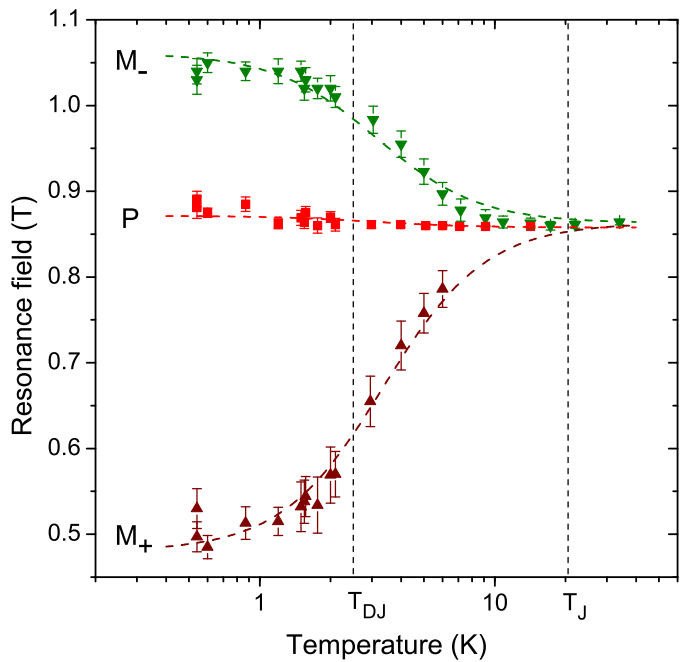}
 \caption{(Color online) Temperature dependence of $27.83$~GHz ESR fields for the components of the doublet $M_{+}$, $M_{-}$
 and of the paramagnetic line $P$, associated with the defects.
 Characteristic temperatures marked on the horizontal axis are $T_{DJ}=\sqrt{DJ}/k_B$ and $T_J=J/k_B.$
 Dashed lines are guide to the eyes.}
 \label{TdepsM1M2@27GHz}
\end{figure}

%Fig4
\begin{figure}[t]
\centering
\includegraphics[width=0.5\textwidth]{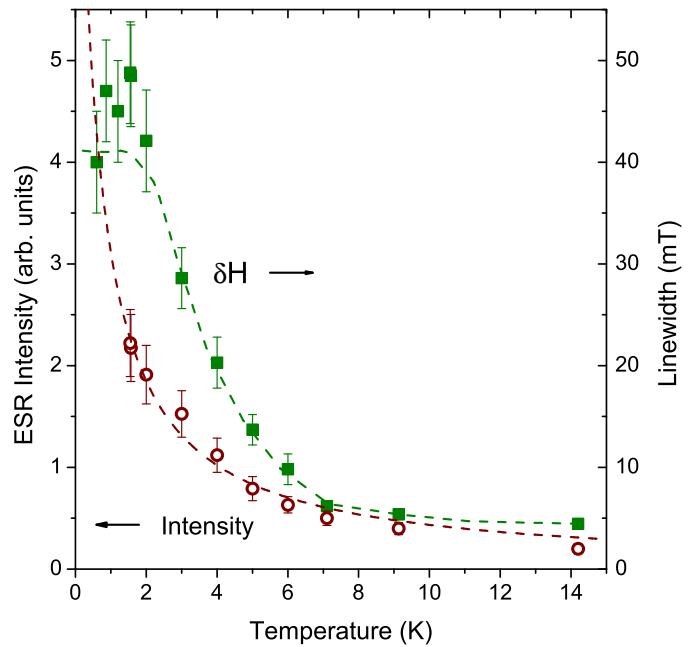}
 \caption{(Color online) Temperature dependence of integral intensity (circles) and
 linewidth (squares, half width at half height)
 of mode $P$ at $\nu=27.83$~GHz. Dashed line passing near squares is guide to the eyes, passing near circles is a $1/T$ Curie-law fit.}
 \label{IntlinewidthP}
\end{figure}

%Fig5
\begin{figure}[t]
\centering
\includegraphics[width=0.5\textwidth]{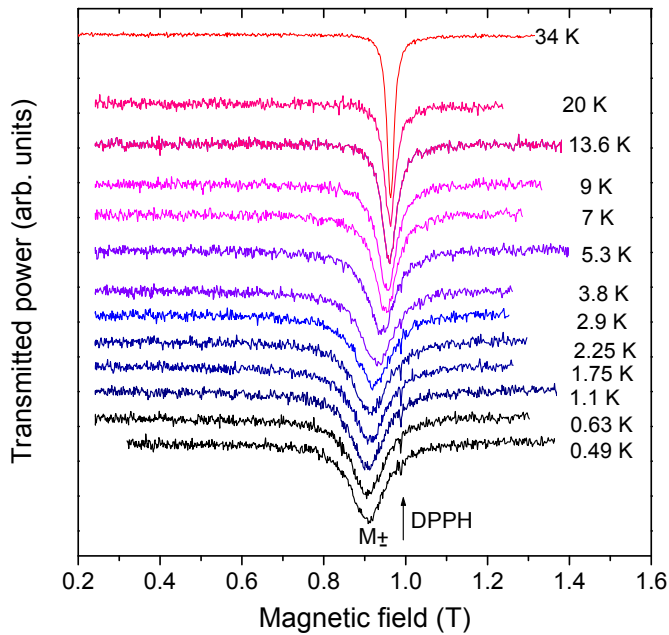}
 \caption{(Color online) Temperature evolution of ESR line at frequency $\nu=27.75$~GHz, ${\bf H} \parallel a$.
  }
 \label{TevolHa27GHz}
\end{figure}

%Fig6
\begin{figure}[t]
\centering
\includegraphics[width=0.5\textwidth]{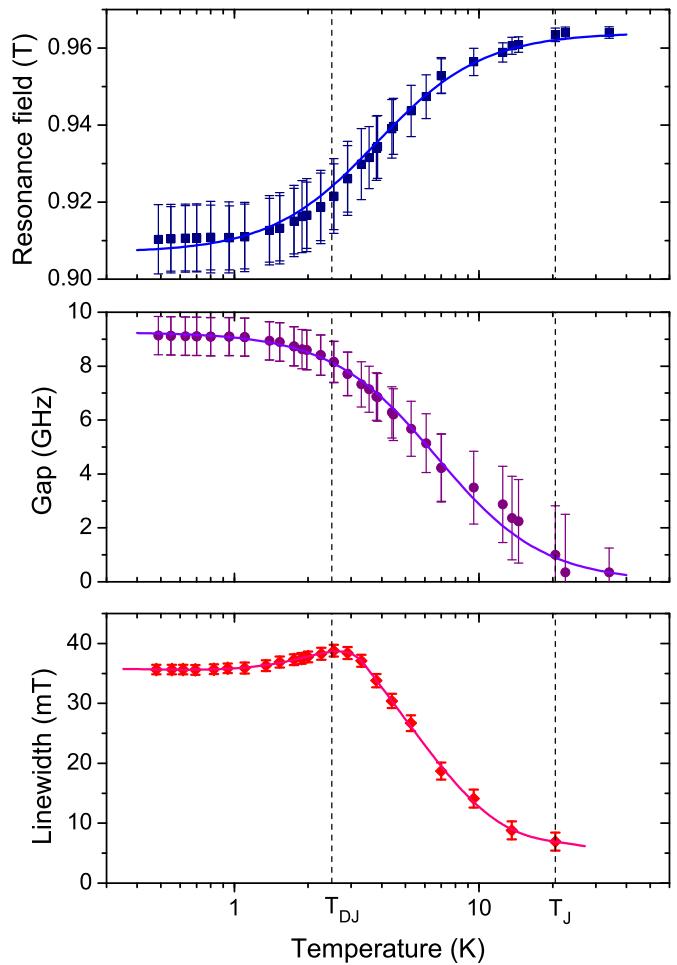}
 \caption{(Color online) Upper panel: Temperature dependence of 27.75~GHz ESR field at ${\bf H} \parallel a$.
 Middle panel: Temperature dependence of the energy gap derived from the resonance field by means of
 relation~(\ref{freqPERP}).
  Lower panel: Temperature dependence of 27.75 GHz ESR linewidth at ${\bf H} \parallel a$.
 Characteristic temperatures marked on the horizontal axis are $T_{DJ}=\sqrt{DJ}/k_B$ and $T_J=J/k_B.$
 Solid lines are guide to the eye.}
 \label{TdepHa27GHz}
\end{figure}

%Fig7
\begin{figure}[t]
\centering
\includegraphics[width=0.5\textwidth]{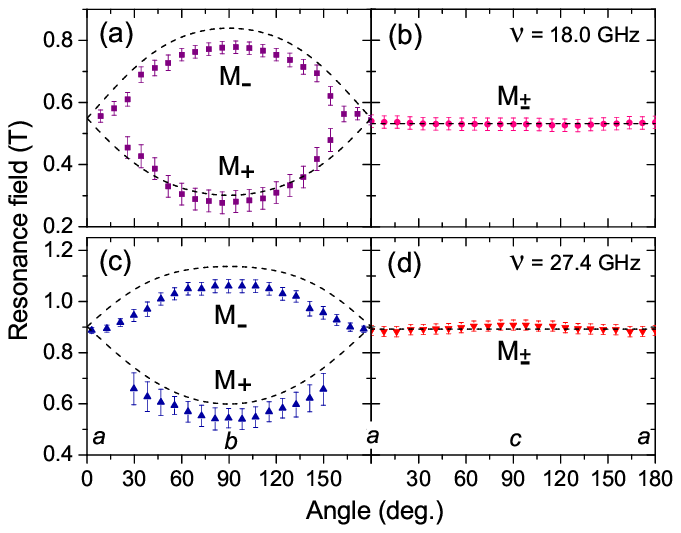}
 \caption{(Color online) Angular dependencies of the resonant
 fields at two frequencies for $T=1.3$~K. 18.0 GHz ESR fields for field in $ab$ plane (upper left panel),
and in $ac$ plane (upper right panel). 27.4 GHz ESR fields for field in $ab$ plane (lower left panel), and in
$ac$ plane (lower right panel)
 Dashed lines are theoretical dependencies according ~(\ref{freqStarykh1}).}
 \label{rota_fields}
\end{figure}

The samples used in our measurements were the single crystals of \KCB\ from the same batch as in
Ref.~\onlinecite{PovarovPRB2014}. We used two  samples with the principal structure axes matching the crystal
facets. A smaller sample A had a mass of 6~mg, and a bigger sample B was about 64~mg. Sample B was used in
experiments in lower frequency range $\nu <$ 25 GHz  where larger size sample is required due to the larger
resonator volume.

\section{Experimental results}\label{Results}

The temperature evolution of the $27.83$~GHz ESR line  is present in Fig.~\ref{Tevol27GHzHb}. Here the magnetic
field is applied along the $b$-axis, parallel to the DM vector ${\bf D}$. We observe  that a single ESR line
with the $g$-factor of $2.24$ found in a paramagnetic phase at $T=30$~K evolves at cooling in a spectrum of
three components, labeled as $M_{+}$, $M_{-}$, $P$. The components $M_{+}$ and $M_{-}$ exhibit a shift from the
paramagnetic resonance field, increasing with cooling. A weak line $P$ remains at the paramagnetic resonance
field.  The temperature dependencies of the resonance fields of modes $M_{+}$ and $M_{-}$, as well as of mode
$P$ are present in Fig.~\ref{TdepsM1M2@27GHz}. The intensity of mode $P$ grows with cooling following Curie law
$1/T$, the  linewidth also increases (see Fig.~\ref{IntlinewidthP}). Due to broadening this line is not so well
visible at most low temperature, as for $T\simeq 4$ K.

%Fig8
\begin{figure}[t]
\centering
\includegraphics[width=0.5\textwidth]{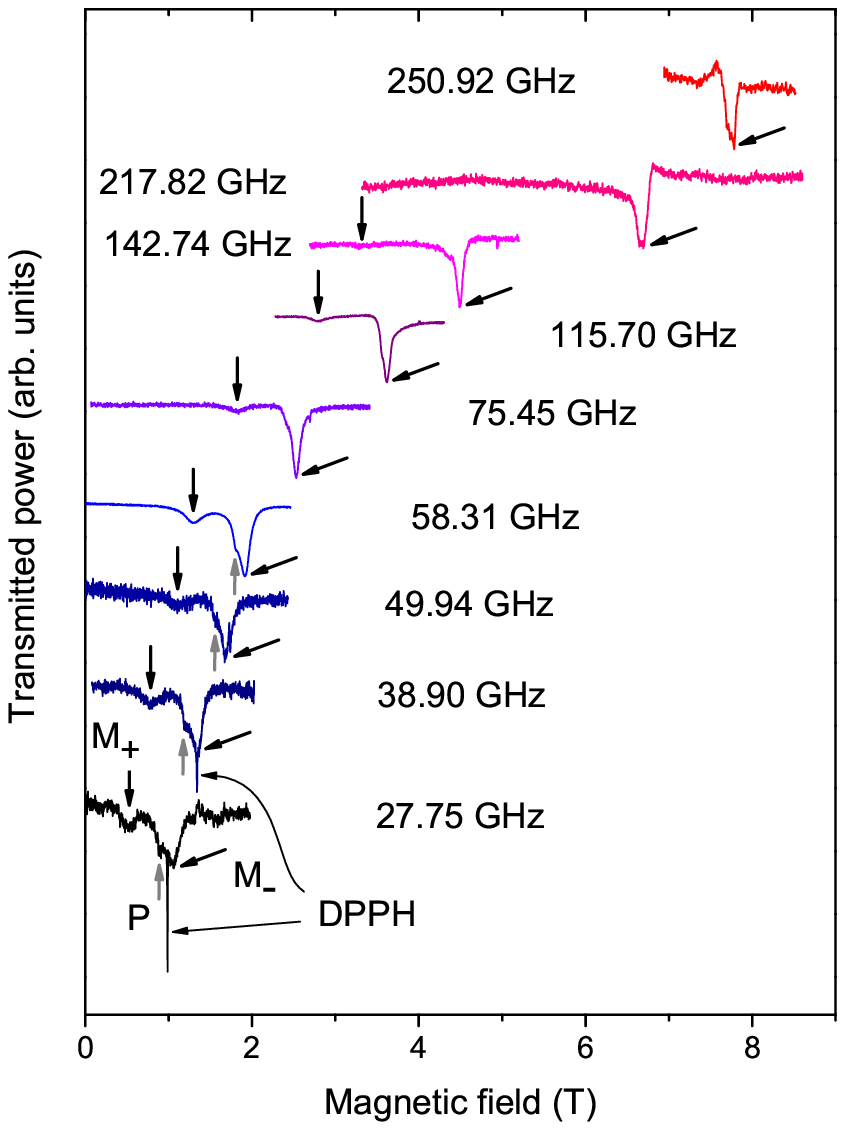}
 \caption{(Color online) Examples of ESR lines at ${\bf H}\parallel b$ for $\nu>25$~GHz, $T=0.5$~K, sample A.
 Black arrows directed down
 mark mode $M_{+}$,
 tilted arrows ---  mode $M_{-}$. Grey arrows directed up mark paramagnetic mode $P$.}
 \label{linesvarfreqHb}
\end{figure}

%Fig9
\begin{figure}[b]
\centering
\includegraphics[width=0.5\textwidth]{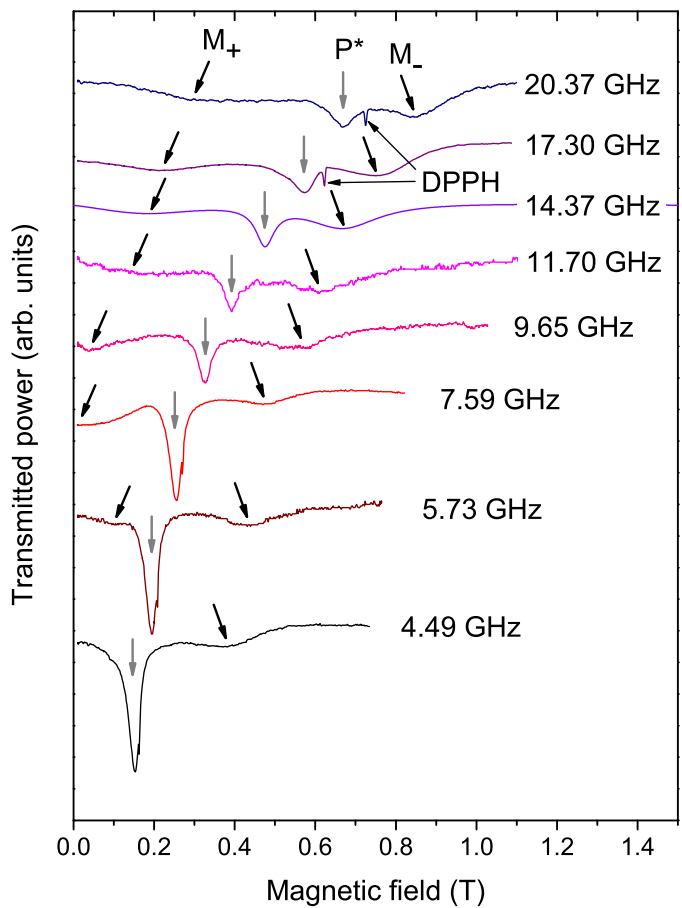}
 \caption{(Color online) Examples of ESR lines at ${\bf H}\parallel b$ for the low frequency range $\nu<25$~GHz,
 $T=1.3$~K, sample B. Right and left tilted black arrows
 mark modes $M_{+}$ and $M_{-}$ correspondingly. Vertical grey arrows mark paramagnetic mode $P^{\ast}$.}
 \label{lineslowfreqHb}
\end{figure}

%Fig10
\begin{figure}[t]
\centering
\includegraphics[width=0.5\textwidth]{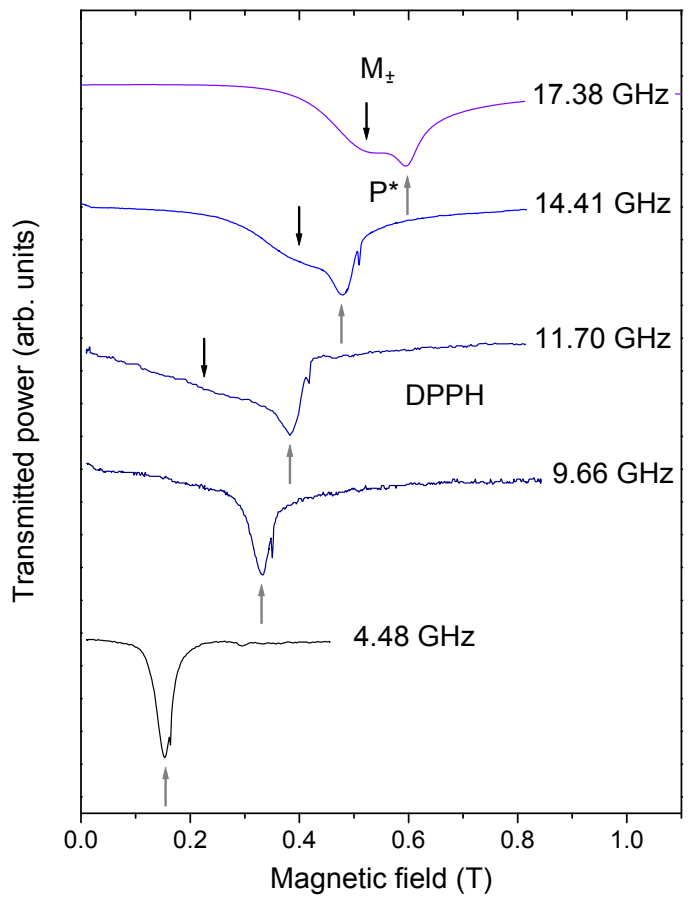}
 \caption{(Color online) Examples of ESR lines at ${\bf H}\parallel a$ for different frequencies at
 $T=1.3$~K, sample B. Black arrows
 indicate mode $M_{\pm}$, grey arrows mark paramagnetic mode $P^{\ast}$.}
 \label{linesvarfreqHa}
\end{figure}

%Fig11
\begin{figure}[t]
\centering
\includegraphics[width=0.5\textwidth]{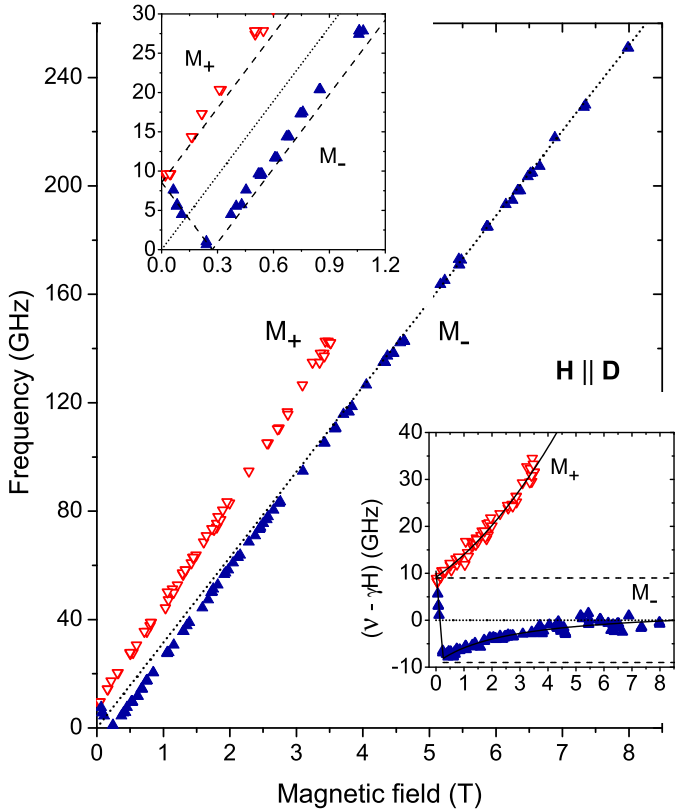}
 \caption{(Color online) Main panel: Frequency-field diagram for ${\bf H} \parallel \mathbf{D}$ at $T=0.5$~K.
 Dotted line presents the  Larmor frequency $\nu_0$~(\ref{Larmor}). Typical error in resonance field does
 not exceed the size of symbols.
 Upper inset: Low-frequency part of the frequency-field diagram. Dashed lines present the theoretical
  prediction~(\ref{freqPAR}) for $D=0.27$~K.
 Lower inset: Resonance shift from the paramagnetic frequency for ${\bf H}
 \parallel b$ at $T=0.5$~K. Dashed lines are drawn according the theoretical prediction~(\ref{freqPAR}) for $D=0.27$~K,
  dotted line marks zero offset from the paramagnetic resonance. Solid lines are guide to the eye.}
 \label{fvsHb}
\end{figure}

%Fig12
\begin{figure}[t]
\centering
\includegraphics[width=0.5\textwidth]{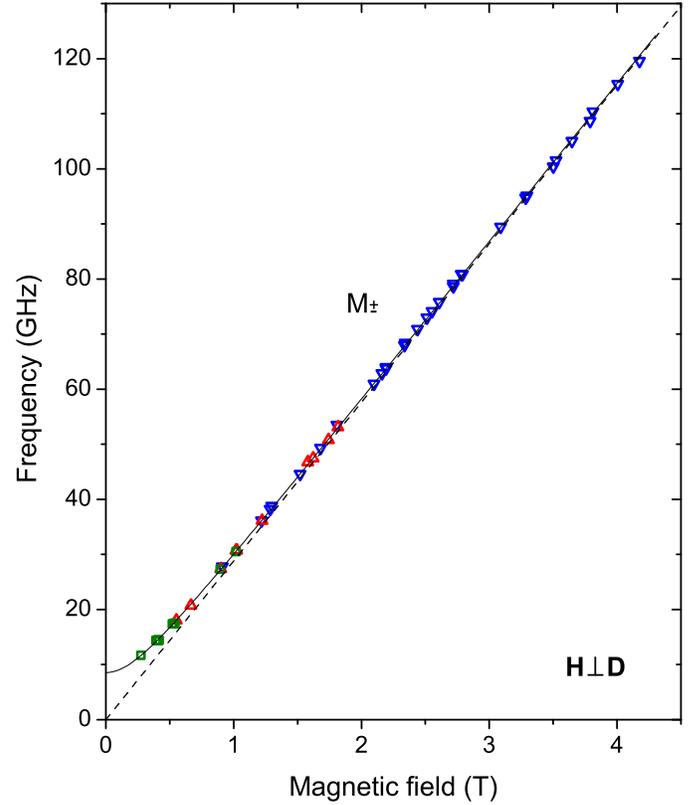}
 \caption{(Color online) Frequency-field diagram for ${\bf H}\perp\mathbf{D}$
 in the low-temperature limit. Data presented by
 $\triangledown$ correspond to sample A at ${\bf H}\parallel a$ and are taken at $T=0.5$~K, $\vartriangle$  present sample B at $T=1.3$~K
 for ${\bf H}\parallel
  c$, $\square$ denote resonance field for sample B at $T=1.3$~K and ${\bf H}\parallel c$.
 Typical error in resonance field does
 not exceed the size of symbols. Solid line presents
 the theory~(\ref{freqPERP}) with parameters $D=0.27$~K, $g_{ac}$=2.06. Dashed line corresponds to ESR
 of the paramagnetic phase~(\ref{Larmor}) at $T>30$~K.}
 \label{fvsHa}
\end{figure}

%Fig13
\begin{figure}[t]
\centering
\includegraphics[width=0.5\textwidth]{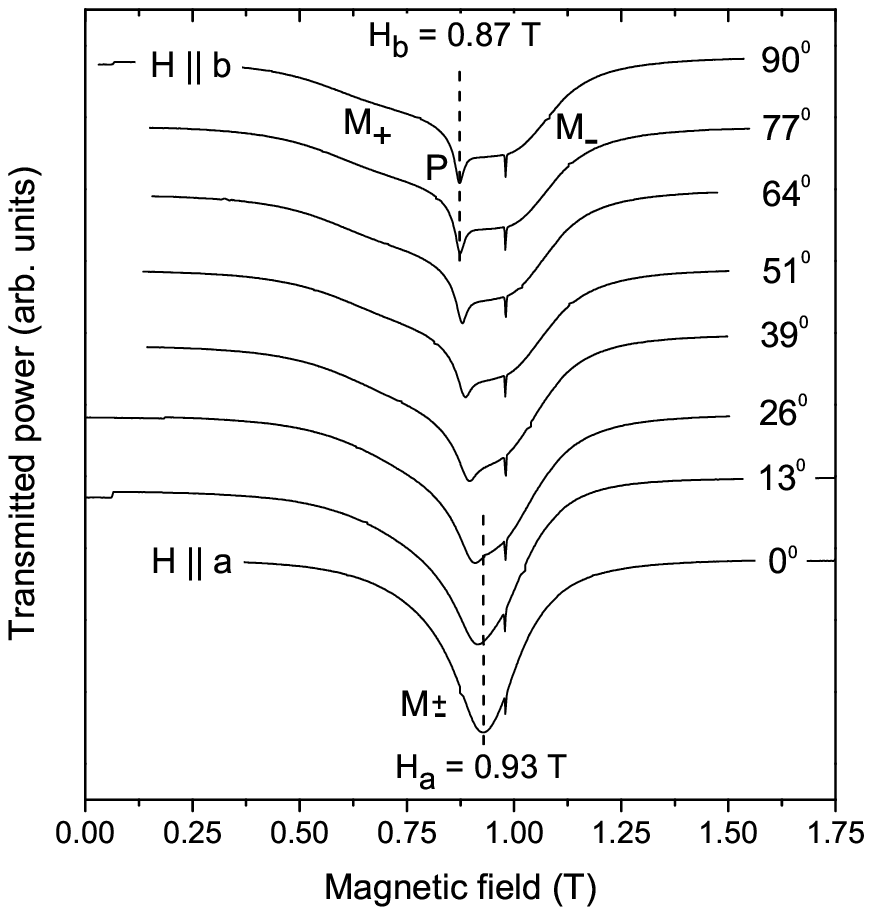}
 \caption{(Color online) Evolution of $\nu=27.44$~GHZ ESR record with the magnetic field rotation in $ab$-plane.
 Sharp resonance line at $H=0.98$~T is the DPPH label with $g=2.0$.}
 \label{rota}
\end{figure}

One can see that the shift of resonance lines $M_{+}$ and $M_{-}$ from the paramagnetic resonance field evolves
with the decrease of temperature and saturates only around $1$~K. These data expand the temperature range of
previous measurements\cite{PovarovPRB2014} to lower temperatures and reveal a saturation of the resonant field
shift. The linewidth of mode $P$ increases with cooling similarly to the increase of the resonant field shift
for modes $M_{-}, M_{+}$. It also saturates at $T\simeq 1$~K.

The temperature evolution of $27.75$~GHz ESR spectrum at ${\bf H}\parallel a$ is present in
Fig.~\ref{TevolHa27GHz}. Here the single line $M_{\pm}$ survives at cooling, but demonstrates a shift to lower
fields. The shift of this single mode also reaches a limit value at $T \simeq 1$~K as shown in
Fig.~\ref{TdepHa27GHz}. The transformation of the single-mode spectrum at ${\bf H}\parallel a$ into a doublet at
${\bf H}\parallel b$ is traced in the angular dependencies shown in Fig.~\ref{rota_fields}. For the rotation of
the magnetic field within the $ac$-plane the single-mode spectrum is conserved and has a negligible angular
dependence, as seen in the right  panels of Fig.~\ref{rota_fields}. The high-temperature ($T=30$~K) $g$-factors
at these orientations $g_a\simeq g_c=2.05\pm 0.01$ are also identical.

 Examples of ESR records
taken at different frequencies for ${\bf H}\parallel a,b$ are presented in Figs.~\ref{linesvarfreqHb},
\ref{lineslowfreqHb}, \ref{linesvarfreqHa}. The data for $\nu>25$~GHz at $T\simeq0.5$~K were taken using the
sample A. For lower frequencies, for the reason described in Sec.\ref{Experiment},  we used sample B, which was
studied at temperatures $T\gtrsim1.3$~K. As the positions of resonance fields  practically do not evolve below
1.3 K, the 1.3 K data present the low-temperature limit values of the resonance fields.  Sample B has a larger
relative intensity of impurity mode ($P$$^*$ in Fig. \ref{linesvarfreqHa}) with respect to the intrinsic signal.
The fraction of paramagnetic defects was estimated from the comparison of the total intensity of ESR signal in
the paramagnetic phase ($T>20$ K), with the intensity of defect ESR measured at $T \simeq 5 K $. Intensity of
the paramagnetic signal at $T>20$ K is ascribed to the total number of spins. The intensities of these two
signals were measured at different temperatures. Therefore we have normalized them to get the ratio of numbers
of spins using the Curie-Weiss law with Curie-Weiss temperature $\theta= J/k_B$  for the main signal at
$k_{B}T>J$ and Curie-Weiss law with $\theta=0$ for ESR of defects. As a result, for sample A the content of
defects was estimated as $x_{A}\simeq0.008$, for sample B $x_{B}\simeq0.02$.

Measurements of resonance fields of the intrinsic signal  at different frequencies are summarized in
frequency-field diagrams in Figs.~\ref{fvsHb}, \ref{fvsHa}. The formation of the zero-field gap of about
$10$~GHz is clearly seen here. Besides,  for ${\bf H}\parallel b$ we see the falling branch, reaching zero
frequency at $H_{0}=0.25$~T and  a rising branch, originating from zero frequency at the same field. The latter
transforms into the component $M_{-}$ of the doublet in the frequency range above the gap. With the increase of
the field the low-field component of the doublet vanishes and above 3.5 T we observe only a single line in the
spectrum.

 For ${\bf
H}\parallel a$ and ${\bf H}\parallel c$ we observe in the whole field range only a single line (mode $M_{\pm}$
in Fig.~\ref{linesvarfreqHa}) originating approximately at $10$~GHz and a paramagnetic line of defects $P$$^*$,
which has different intensity for different samples. At frequencies lower than $10$~GHz we observe only the
paramagnetic ESR line of defects.

The low-intensity residual paramagnetic line P correlates with the main ESR signal. At first, the broadening of
mode $P$ at cooling below $4$~K has a temperature dependence analogous to that of the resonance field shifts for
modes $M_{+}$, $M_{-}$ and $M_{\pm}$. The second evidence for correlation may be found in the angular dependence
of ESR response taken at $T=4.2$ K, which is present in Fig.~\ref{rota}. Here we see a narrow paramagnetic
resonance line with $g\simeq2.24$ at ${\bf H}\parallel b$. With the rotation of the magnetic field towards the
$a$-axis this narrow line shifts towards $M_{-}$ and then merges with the main signal. As a result, at ${\bf H}
\parallel a$ there is a single line with $g\simeq2.10$. If the narrow line $P$ would have survived as an independent signal,
it would be seen as a narrow sharp peak on the background of a wider
resonance line.

\section{Discussion}
\subsection{ESR modes of a spin chain}

The spinon continuum and the related ESR doublet observed in \KCB\ are consequences of quantum fluctuations in
quasi-1D spin $S=1/2$ system, which remains disordered far below the Curie--Weiss temperature. The observed
formation of the ESR doublet $M_{+}$, $M_{-}$ in the temperature range below the temperature $T_J=J/k_{B}$ is
fully consistent with the scenario of the spin chain entering a quantum critical regime upon cooling. The
anisotropic behavior of the doublet is in a perfect agreement with the symmetry considerations for
Dzyaloshinskii--Moriya vector ${\bf D}$. The symmetry of the orthorhombic $Z=4$ Pnma structure of \KCB\ implies
${\bf D}\parallel b$, as shown in Fig.~\ref{structure}. The theory prediction
(Refs.~\onlinecite{StarykhESR,Povarov2011}) is that the ESR doublet should be observable only when the magnetic
field has a component along ${\bf D}$. In a full correspondence with the above symmetry restrictions we observe
a doublet at ${\bf H}
\parallel b$ and a single line at ${\bf H}
\perp b$.  For an arbitrarily aligned magnetic field ${\bf H} = (H_a, H_b, H_c)$ and ${\bf D}
\parallel b$ the theoretical analysis\cite{Povarov2011} predicts two following resonance frequencies:

\begin{equation}
\begin{aligned}
 (2 \pi \hbar \nu_{+})^2  =&    (g_{a}\mu_{B}H_{a})^{2} +  (g_{b}\mu_{B}H_b + \pi D/2)^{2}+\\
 &(g_{c}\mu_{B}H_c )^{2},\\
(2 \pi \hbar \nu_{-})^2  =&   (g_{a}\mu_{B}H_{a})^{2} +  (g_{b}\mu_{B}H_b -  \pi D/2)^{2}+\\
 &(g_{c}\mu_{B}H_c )^{2}.
    \label{freqStarykh1}
    \end{aligned}
\end{equation}

In the limiting cases of $\mathbf{H}\parallel\mathbf{D}$ and $\mathbf{H}\perp\mathbf{D}$ the equation
set~(\ref{freqStarykh1}) correspondingly transforms into:

\begin{eqnarray}
 &2&\pi\hbar\nu_{\pm}=\left|g_{b}\mu_{B}H \pm \dfrac{\pi D}{2}\right|,
 \label{freqPAR}
\\
&2&\pi\hbar\nu_{\pm}=\sqrt{(g_{\perp}\mu_{B}H)^{2} +  \left(\dfrac{\pi D}{2}\right)^{2}}. \label{freqPERP}
\end{eqnarray}

Here $g_{\perp}$ is the value of g-factor for the field oriented perpendicular to $b$. These relations give the
zero-field gap $2\pi \hbar \Delta=\pi D/2$ and the soft mode for $\mathbf{H}\parallel\mathbf{D}$ is predicted to
occur at $H_{0}=\pi D/(2g_{b}\mu_{B})$.

\subsection{Data analysis}

In the above description of the low-temperature spectrum $D$ is the \emph{only} open parameter, as the
$g$-factor values are given by the paramagnetic resonance at high temperatures (when $k_{\text{B}}T\gtrsim J$).
This single parameter is easily determined from the zero-field energy gap $\Delta$. We derive  $\Delta=8.7$~GHz
by fitting the experimental $\nu(H)$ dependence for ${\bf H} \parallel a$ (Fig.\ref{fvsHa}) with  the relation
(\ref{freqPERP}). Thus we get the value of $D=0.27\pm 0.015$~K. The frequency--field dependencies calculated for
${\bf H}\parallel b$ according to equation (\ref{freqPAR}) are shown in the upper and lower inserts of
Fig.~\ref{fvsHb} by dashed lines. Both rising and falling branches, as well a change from falling to rising
behavior at $H_{0}=0.25$~T are in a good agreement with the theoretical predictions  in the low-field regime at
$H<$1.2 T. We would like to note that observation of such a soft mode at $H_{0}$ is impossible in the other
model compound \ccc\ due to non-collinear arrangement of DM vectors in adjacent chains. The low-frequency
angular dependence of the ESR fields, taken at the rotation of the magnetic field in $ab$- and $ac$-planes is
also in a good agreement with the theoretical relations (\ref{freqStarykh1}), as shown in the upper panel of
Fig.~\ref{rota_fields}.

Nevertheless, the agreement gets worse with the field increase. Positions of the doublet components start to
deviate from the calculated frequencies. One can see this behaviour in frequency--filed diagrams of
Fig.~\ref{fvsHb} as well as in $\nu\simeq27$~GHz angular dependence in the lower panel of
Fig.~\ref{rota_fields}. The resonance shift of the component $M_{+}$ rises in a non-linear way with the
increasing field, while for the component $M_{-}$ it gradually disappears. This behavior is emphasized in the
``resonance shift'' representation shown in the lower inset of Fig.~\ref{fvsHb}. As the mode $M_{+}$ vanishes
around $4$~T, even a qualitative agreement with the theory predictions is completely lost. The spectrum becomes
indistinguishable from a conventional paramagnet
--- a single line at Larmor frequency:

\begin{equation}
2 \pi \hbar \nu_0 = g_b\mu_{B}H. \label{Larmor}
\end{equation}

For $\mathbf{H}\perp\mathbf{D}$ the resonance at such high frequencies becomes indistinguishable from a
paramagnetic spectrum as well, being, nevertheless, in accordance with theoretical relation (\ref{freqPERP}).

On the one hand, this transformation to the Larmor-type ESR spectrum is not unexpected as the magnetic field
suppresses quantum fluctuations which are the natural ground for the spinon continuum, and hence for the spinon
ESR doublet. The theoretical analysis of the spinon continuum in a magnetic field predicts a collapse of the
continuum at the saturation field.\cite{Mueller} Besides,  the appearance of the spinon doublet in ESR response
is related to the effect of \emph{weak} DM interaction. The actual observed field of the doublet collapse is
only $3.5$~T, constituting about $0.13H_{\text{sat}}$ (the saturation field of \KCB\ is about 27 T
\cite{PovarovPRB2014}). Such a low value of the collapse field indicates that the corresponding energy scale is
related to the DM interaction. One can empirically estimate the collapse field as $\sqrt{H_{\text{sat}}
D/(g\mu_{B}) }\simeq\sqrt{2DJ}/(g\mu_{B})\simeq3$~T. For the previously studied material \ccc\ the doublet
collapse was observed at the magnetic field  of about $4$~T,\cite{SmirnovPRB2015} which is a half of the
saturation field. However, the DM interaction strength in this compound is larger, corresponding to zero-field
ESR gap of $14$~GHz. The estimation according to the above empiric rule gives $4$~T for \ccc\ in a good
agreement with the experiment. It is worth to note, that the final formation of the doublet and the maximum of
the resonance field shift for the doublet components in \KCB\ are achieved at $T^{*}\simeq1$~K, which is much
lower, than the characteristic exchange temperature $J/k_{\text{B}}\simeq20$~K. The value of $T^*$, analogous to
the field of the doublet collapse is rather of the order of intermediate temperature $T_{DJ}=\sqrt{DJ}/k_{B}$ ,
than of the order of the exchange temperature $T_J$.

\subsection{Paramagnetic defects}

The paramagnetic mode $P$ observed at ${\bf H} \parallel b$ in sample A may be separated from the ESR signal of
the main matrix in the temperature interval $1.3<T<14$~K due to its narrow line and the position, different from
the resonances $M_{-}$ and $M_{+}$  (see Fig.~\ref{Tevol27GHzHb}). The position of line $P$ does not depend on
temperature and corresponds to the $g$-factor value $g_{b}$ of the paramagnetic phase within the experimental
accuracy. This indicates that the mode $P$ may originate from intrinsic defects such as breaks of the spin
chains. The growth of the linewidth of signal $P$ at cooling is similar to the growth of the energy gap $\Delta$
and the mode $M_{\pm}$ linewidth, as shown in Fig.~\ref{TdepHa27GHz}. This observation confirms the proposition
on the interaction between the defects and the spin chains.

 The growth of the linewidth may be qualitatively explained considering the effective exchange
interaction between the defects and spin chains using a concept of the exchange narrowing and broadening
developed by Anderson.\cite{PWAnderson} Indeed, the exchange interaction $J^*$ between two magnetic systems with
the resonance frequencies differing for $\delta \omega$ is known to modify the ESR spectrum. In case of the
``slow'' exchange ($J^*/\hbar<\delta \omega$), the lines of the initial spectrum are broadened, while at $\delta
\omega = 0$ the broadening is absent. This may presumably explain the broadening of the mode $P$ at cooling as
the difference in resonance frequencies of modes $M_{-}$, $M_{+}$ and mode $P$ develops. In case of a ``fast''
exchange ($J^*/\hbar>\delta \omega$) the ESR lines should merge into a single narrow line.\cite{PWAnderson} The
vanishing of the defect mode $P$ at the rotation of the magnetic field in the $ab$-plane, present in
Fig.~\ref{rota} may be attributed to the transition from the ``slow'' exchange at ${\bf H}\parallel b$ (lines
are observed separately) to the ``fast'' exchange at ${\bf H}\parallel a$, when the difference between resonance
frequencies becomes smaller because of the angular dependence and the lines may merge due to the effective
exchange $J^*$. In the absence of interaction between the defects and spin-chain matrix the narrow line $P$
should be observable together with the wider line $M_{-}$.

\section{Conclusion}

A detailed ESR study of the quasi-1D $S=1/2$  antiferromagnetic system with uniform Dzyaloshinskii-Moriya
interaction was performed. The energy gap of $8.7$~GHz was observed in zero field and a characteristic doublet
of resonance lines was found in a magnetic field. The zero-field gap, ESR doublet and its anisotropy as well as
the frequency-field dependencies for three orientations of the magnetic field are in a good agreement with the
theory of spinon continuum modification by uniform Dzyaloshinskii--Moriya interaction, compatible with crystal
symmetry of \KCB. There is only one fitting parameter in this theory. On the other hand, with the increase of
the magnetic field the  low-frequency approximation fails, and the deviation from the linear frequency-field
dependencies in a high field and vanishing of the doublet in a rather low field of $0.13H_{\text{sat}}$ are not
understood now. The system of magnetic defects, interacting with the spin excitation of the main matrix was also
revealed by ESR experiment.

\section{Acknowledgements} \label{Acknowledgements}
  We thank   V.N.~Glazkov, S.S.~Sosin,
L.E.~Svistov for numerous discussions and comments, and for help
with microwave spectrometers, O.A.~Starykh for discussions. Work at
the Kapitza Institute is supported by Russian Foundation for Basic
Research, grant No. 15-02-05918, and by the Program for Basic
Research of the Presidium of Russian Academy of Sciences.
 ETHZ team acknowledges support from Swiss National
Science Foundation, Division~2.

\bibliography{BibfileESRKCuSOBr}

\end{document}